\documentclass[trackchanges,twocolumn]{aastex7}
\usepackage{color}

\begin{document}

\title{The Origin of the Mg-rich Supernova Remnant J0550–6823 and the Frequency of Similar Events in the Large Magellanic Cloud}

\author[0009-0000-8310-9463]{Yui Kuboike}
\affiliation{Meiji University, Department of Physics}
\email{ce257009@meiji.ac.jp}

\author[0000-0001-9267-1693]{Toshiki Sato}
\affiliation{Meiji University, Department of Physics}
\email{toshiki@meiji.ac.jp}

\author[0000-0002-8152-6172]{Hiromasa Suzuki}
\affiliation{University of Miyazaki,Faculty of Engineering}
\email{suzuki@astro.miyazaki-u.ac.jp}

\author[0009-0003-0653-2913]{Kai Matsunaga}
\affiliation{Kyoto University, Department of Physics}
\email{matsunaga.kai.i47@kyoto-u.jp}

\author[0000-0003-1518-2188]{Hiroyuki Uchida}
\affiliation{Kyoto University, Department of Physics}
\email{uchida@cr.scphys.kyoto-u.ac.jp}

\author[0000-0002-8816-6800]{John P. Hughes}
\affiliation{Rutgers University, Department of Physics and Astronomy}
\email{jph@physics.rutgers.edu}

\author[0000-0003-1415-5823]{Paul P. Plucinsky}
\affiliation{Center for Astrophysics — Harvard \& Smithsonian}
\email{pplucinsky@cfa.harvard.edu}

\begin{abstract}
Shell burning and internal mixing in massive stars play an important role in setting the initial conditions for core-collapse supernova explosions. In the late stages of stellar evolution, intense shell burning can cause distinct convective regions to merge, fundamentally restructuring the stellar interior. Although such phenomena are difficult to observe directly, the observation of ``Mg-rich'' supernova remnants (SNRs) has recently emerged as a potential signature of these events. In this study, we reanalyze X-ray observations of J0550--6823, a SNR in the Large Magellanic Cloud (LMC) and a new candidate Mg-rich SNR. Our spectral analysis confirms a low Ne/Mg mass ratio of $\approx$1, and its classification as Mg-rich. By comparing the observational results with pre-supernova models, we suggest that the progenitor of J0550-6823 likely had an extended convective shell that reduces the Ne/Mg ratio prior to its explosion. Furthermore, we observe that $\sim$2--3 Mg-rich SNRs exist in the LMC, suggesting that $\lesssim$10--40\% of massive stars in the LMC may have had an extended convective shell, similar to what we observed in J0550-6823. This fraction would be important for understanding the final stages of the evolution of massive stars and galactic chemical evolution.
\end{abstract}


\section{introduction}
\label{sec:intro}
Shell burning processes in the final days before core collapse play a crucial role in shaping the internal structure of massive stars \citep[e.g.,][]{2002RvMP...74.1015W,2014ApJ...783...10S}, which is thought to significantly affect the initial conditions of the supernova explosion \citep[e.g.,][]{2016MNRAS.460..742M,2016ApJ...816...43S,2016ApJ...818..124E} and the chemical composition of supernovae (SNe) and their remnants \citep{2020A&A...642A..33D,2021A&A...652A..64D,2024ApJ...970....4M,2025ApJ...984..185S,2025ApJ...990..103S}. Among the late-stage evolution, the so-called “shell merger”, in which an inner O-burning shell dynamically mixes with an overlying C-burning shell, has recently garnered significant attention  \citep[e.g.,][]{2011ApJ...733...78A,2020ApJ...890...94Y,2021A&A...656A..58L,2024MNRAS.533..687R}.

\begin{figure*}[t!]
  \centering
  \gridline{
    \includegraphics[bb=0 0 2600 1080,width=0.95\linewidth]{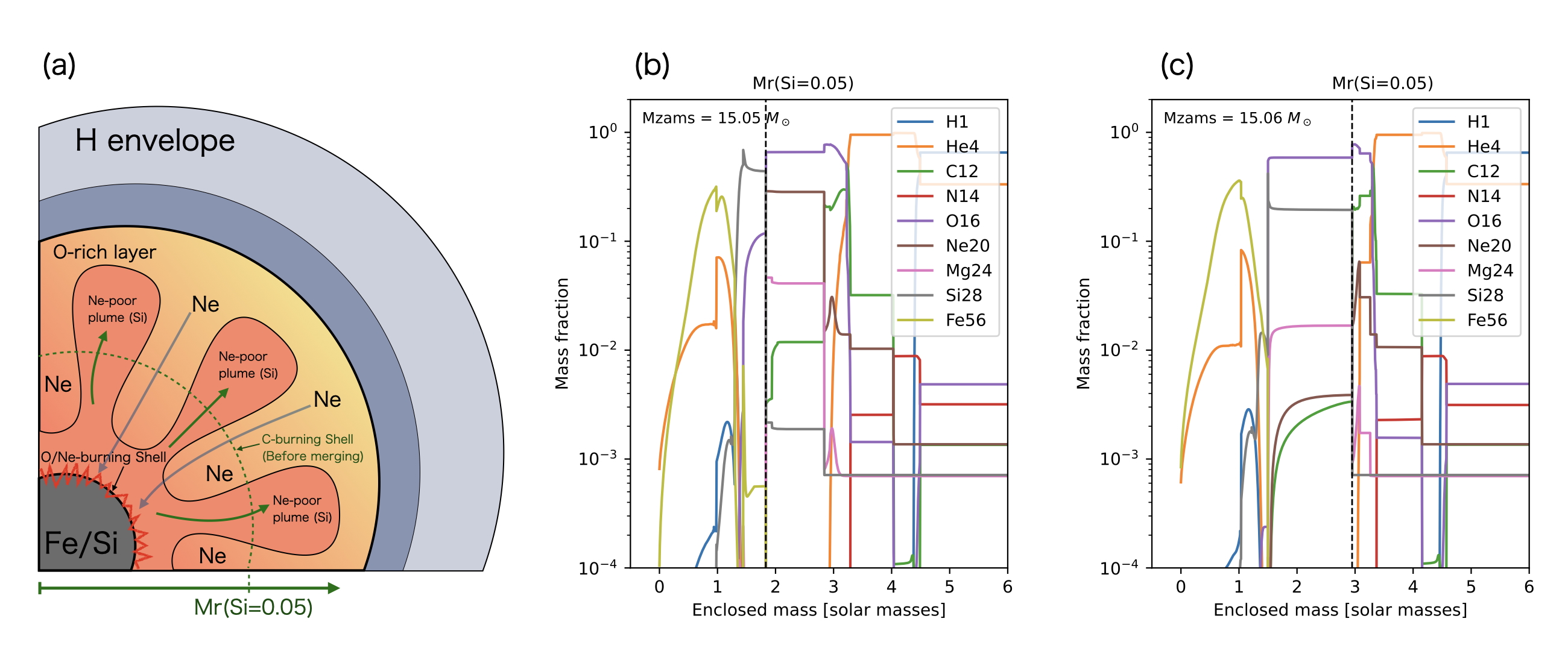}}
  \vspace{-7mm}
  \caption{
    (a) The internal structure of a star, which experiences an active shell burning process.
    (b) Mass fraction profile of the 15.05 $M_{\odot}$ model. The Si mass fraction that reaches 0.05 is shown by the dashed line.
    (c) Same as (b), but for the 15.06 $M_{\odot}$ model. 
    The models are taken from \citet{2018ApJ...860...93S}.}
  \label{shellmerger}
\end{figure*}

As illustrated in Figure \ref{shellmerger} (a), Si synthesized in the O-burning shell can be transported outward into the C/Ne-burning shell, leading to a substantial increase in the Si mass fraction (Figures 1(b) and (c)). Shell mergers can markedly alter the progenitor’s structure and composition at the moment of explosion. Not limited to shell mergers, similar extended convective shells in the final stages have also been suggested in scenarios such as “Ne-shell intrusion" \citep{2025ApJ...984..185S}. In the Ne-shell intrusion case, the Ne-burning shell merges into the outer layer. At this moment, the outer shell is not a burning shell. Notably, three-dimensional hydrodynamic simulations suggest that such vigorous convective shells may promote successful supernova explosions by modifying the core compactness \citep{2025ApJ...984..185S,2025A&A...695A..71L} and seeding perturbations in the stellar interior \citep{2013ApJ...778L...7C,2021ApJ...915...28B}. Despite their theoretical implications, direct observational evidence for such active shell burning process remains scarce.

Recently, ``Mg‐rich'' Supernova Remnants (SNRs) have been proposed as potential remnants of progenitors that experienced shell mergers \citep{2025ApJ...984..185S,2024ApJ...970....4M}. In most SNRs, the Mg abundance is substantially lower than that of Ne; for example, the solar Mg/Ne mass ratio is 0.37--0.56 \citep{1989GeCoA..53..197A,2009ARA&A..47..481A}. By contrast, Mg‐rich SNRs exhibit Mg/Ne ratios near unity or higher \citep[e.g.,][]{2003ApJ...592L..41P,2017ApJ...834..189P}, classifying them as Mg‐rich (or Ne‐poor). This enhancement is naturally explained if active late-stage shell processes  efficiently burn the ingested Ne, thereby increasing the Mg/Ne ratio \citep{2024ApJ...970....4M,2025ApJ...984..185S}. The Large Magellanic Cloud (LMC) remnant N49B was the first identified Mg‐rich SNR in the Magellanic Clouds \citep{2003ApJ...592L..41P}, and recent work has linked its ejecta composition to a shell-merger or similar event \citep{2025ApJ...984..185S}. Accordingly, systematic observations of Mg‐rich SNRs offer a promising avenue to probe extended shell convection activity and the late‐stage evolution of massive stars.

J0550--6823 is a relatively old SNR in the LMC, previously investigated by \citet{2023ApJ...950...74S}. Its outer shell exhibits an elliptical morphology with angular dimensions of $4\farcm9 \times 6\farcm1$, ranking it among the largest known remnants in the LMC (total size $\sim71\times88 
\,\rm{pc}$). X-ray spectra reveal strong emission lines from O, Ne, Mg, and Fe, and modeling of these lines indicates a Type II explosion from a progenitor star exceeding $20\,M_\odot$ \citep{2023ApJ...950...74S}. In particular, its Mg‐rich composition stands out as one of its important characteristics. Thus, this remnant could be the second candidate as an Mg-rich SNR in the LMC and is important for understanding the frequency of stellar activity such as Ne-shell intrusion or shell merger. However, the remnant’s faint X-ray emission makes its spectrum highly sensitive to background‐subtraction uncertainties, complicating robust abundance determinations. Consequently, the designation of J0550--6823 as an Mg‐rich SNR remains tentative. Moreover, no direct comparison with pre‐supernova models has yet been undertaken, and the potential connection to late‐stage stellar processes is still unresolved.

In this study, we use an improved background modeling tool to robustly confirm the elemental abundances of J0550--6823 and investigate the possibility of a pre‐supernova shell-merger event for this remnant. We also estimate the Mg-rich fraction of supernovae in the LMC that may have undergone active shell burning process such as shell mergers and discuss the implications for galactic chemical evolution. \citet{2018MNRAS.474L...1R} demonstrated that shell mergers can dramatically enhance the production of odd‐Z elements such as K, Sc, and Cl, which are otherwise underproduced in standard stellar evolution. If about 50$\%$ of massive stars experience a shell merger, this process could resolve the long‐standing discrepancy between observed odd‐Z abundances and theoretical predictions \citep{2006ApJ...653.1145K,2020ApJ...900..179K,2018MNRAS.476.3432P}. Consequently, quantitatively determining the fraction of LMC SNRs that have undergone shell mergers is essential for understanding the chemical enrichment history of the LMC’s interstellar medium.

\section{Observation and spectrum fitting of J0550--6823} \label{sec:observation}

\begin{figure}[t!]
    \centering
    \vspace{5mm}
    \includegraphics[bb=0 0 1742 1647,width=0.8\linewidth]{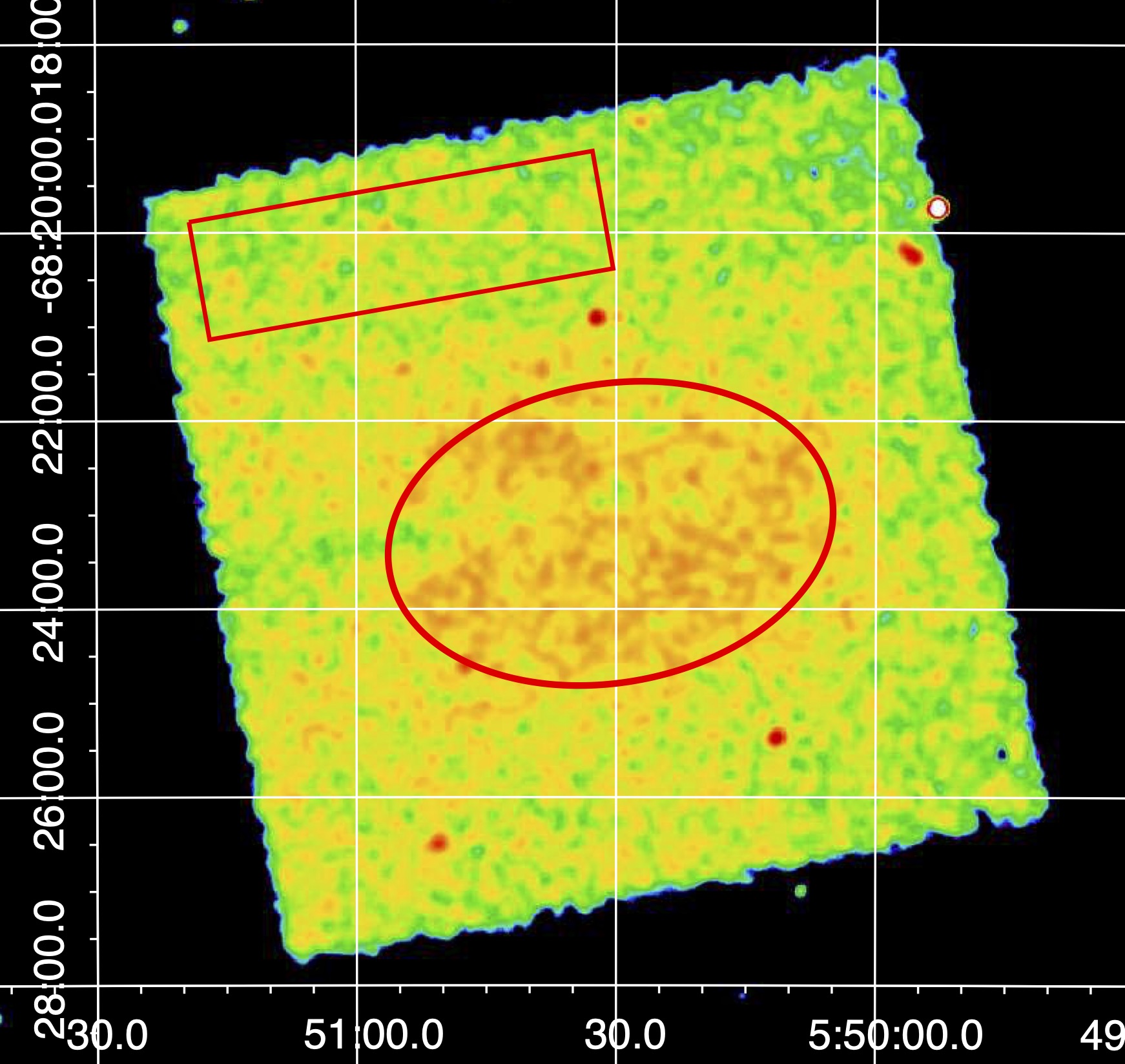}
    \caption{Chandra X-ray image of J0550--6823. The red ellipse shows the outer shell, which was used as the extraction region for spectral analysis. The rectangle region is used for modeling sky background.}
    \label{J0550region2}
\end{figure}

Chandra ACIS has observed the LMC SNR J0550--6823 on 2003 October 2 with a total exposure of 67.48 ks. This paper employs a list of Chandra datasets, obtained by the Chandra X-ray Observatory, contained in~\dataset[DOI:10.25574/cdc.472]{https://doi.org/10.25574/cdc.472}.
We here reprocessed the event file of this observation using CIAO version 4.16 and CalDB 4.11.2. Figure \ref{J0550region2} shows the Chandra X-ray image of J0550--6823 using 0.5--10 keV. We extracted the X-ray spectrum from the ellipse red region in the figure. As explained in Section \ref{sec:intro}, it is difficult to visualize the entire J0550--6823 region due to its weak X-ray emission from the remnant, making its boundaries ambiguous. 

\begin{figure}[t!]
    \centering
    \includegraphics[bb=0 0 2750 2125,width=\linewidth]{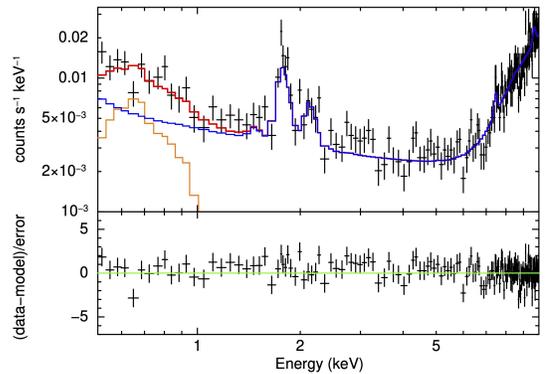}
    \caption{X-ray spectrum extracted from the rectangle region in Fig.~\ref{J0550region2} to model the sky background. The orange line represents the plasma model for the foreground emission. The blue line represents the background spectrum generated by {\tt mkacispback}. The red line represents the final result of the background model with photon absorption model, foreground emission model ($\rm phabs\times apec$), and the acispback model.}
    \label{background2}
\end{figure}

\begin{table*}[htbp]
 \centering
       \caption{The best-fit parameters of the spectral fittings.}
   \label{taubestfit_transpose}
  
   \hspace{-20mm}
   \begin{tabular}{lcccc} \hline
    $\tau~[\rm s \cdot cm^{-3}]$ & $(a)~5.00 \times 10^{10}$ & $(b)~1.00 \times 10^{11}$ & $(c)~5.00 \times 10^{11}$ & $(d)~1.00 \times 10^{12}$ \\ \hline
    $k_e T$ [keV] & $0.51\pm0.01$ & $0.41\pm0.01$ & $0.32\pm0.01$ & $0.30\pm0.01$  \\ 
    (O/H)/(O/H)$_\odot$ & $0.37\pm0.01$ & $0.25\pm0.01$ & $0.45^{+0.05}_{-0.07}$ & $0.63^{+0.36}_{-0.06}$ \\ 
    (Ne/H)/(Ne/H)$_\odot$ & $0.58\pm0.02$ & $0.42\pm0.02$ & $0.51^{+0.02}_{-0.06}$ & $0.63^{+0.23}_{-0.05}$\\ 
    (Mg/H)/(Mg/H)$_\odot$ & $1.12^{+0.07}_{-0.12}$ & $0.92\pm0.06$ & $1.18^{+0.36}_{-0.09}$ & $1.58^{+0.83}_{-0.27}$ \\ 
    (Si/H)/(Si/H)$_\odot$ & $0.64\pm0.19$ & $0.63^{+0.19}_{-0.18}$ & $1.21^{+0.32}_{-0.30}$ & $1.77^{+0.55}_{-0.45}$\\ 
    (Fe/H)/(Fe/H)$_\odot$ & $0.15^{+0.01}_{-0.02}$ & $0.10\pm0.01$ & $0.11^{+0.03}_{-0.01}$ & $0.14^{+0.07}_{-0.03}$ \\
    Redshift [$10^{-3}$] & $1.90^{+0.02}_{-0.20}$ & $1.85^{+0.03}_{-0.22}$ & $8.13^{+0.25}_{-0.33}$ & $8.26^{+0.24}_{-0.37}$  \\ 
    norm [$\frac{10^{-18}}{4\pi D^2}\int n_en_HdV\, \text{cm}^{-5}$] & $3.10^{+0.27}_{-0.14}$ & $6.16^{+0.08}_{-0.10}$ & $9.20^{+0.11}_{-0.10}$ & $8.68^{+1.17}_{-2.41}$ \\ 
    $\chi^{2}/\rm d.o.f$ & 464.42/384 & 462.68/384 & 449.55/384 & 444.71/384  \\\hline
    Mass ratios$^{*}$&\\
    $M_{\rm Ne}$/$M_{\rm Mg}$ & $1.40_{-0.16}^{+0.11}$ & $1.22\pm0.10$ & $1.15_{-0.17}^{+0.36}$ & $1.07_{-0.19}^{+0.69}$   \\
    $M_{\rm Si}$/$M_{\rm Mg}$ & $0.62_{-0.20}^{+0.19}$ & $0.73_{-0.22}^{+0.23}$ & $1.10_{-0.29}^{+0.41}$ & $1.21_{-0.37}^{+0.73}$   \\
    \hline
   \end{tabular}

  \tablenotetext{*}{The solar values of \cite{1989GeCoA..53..197A} were used to estimate Si/Mg and Ne/Mg mass ratios.}

\end{table*}

\begin{figure}[!htbp]  
  \centering
          \includegraphics[bb=0 0 2256 1569,width=1.0\linewidth]{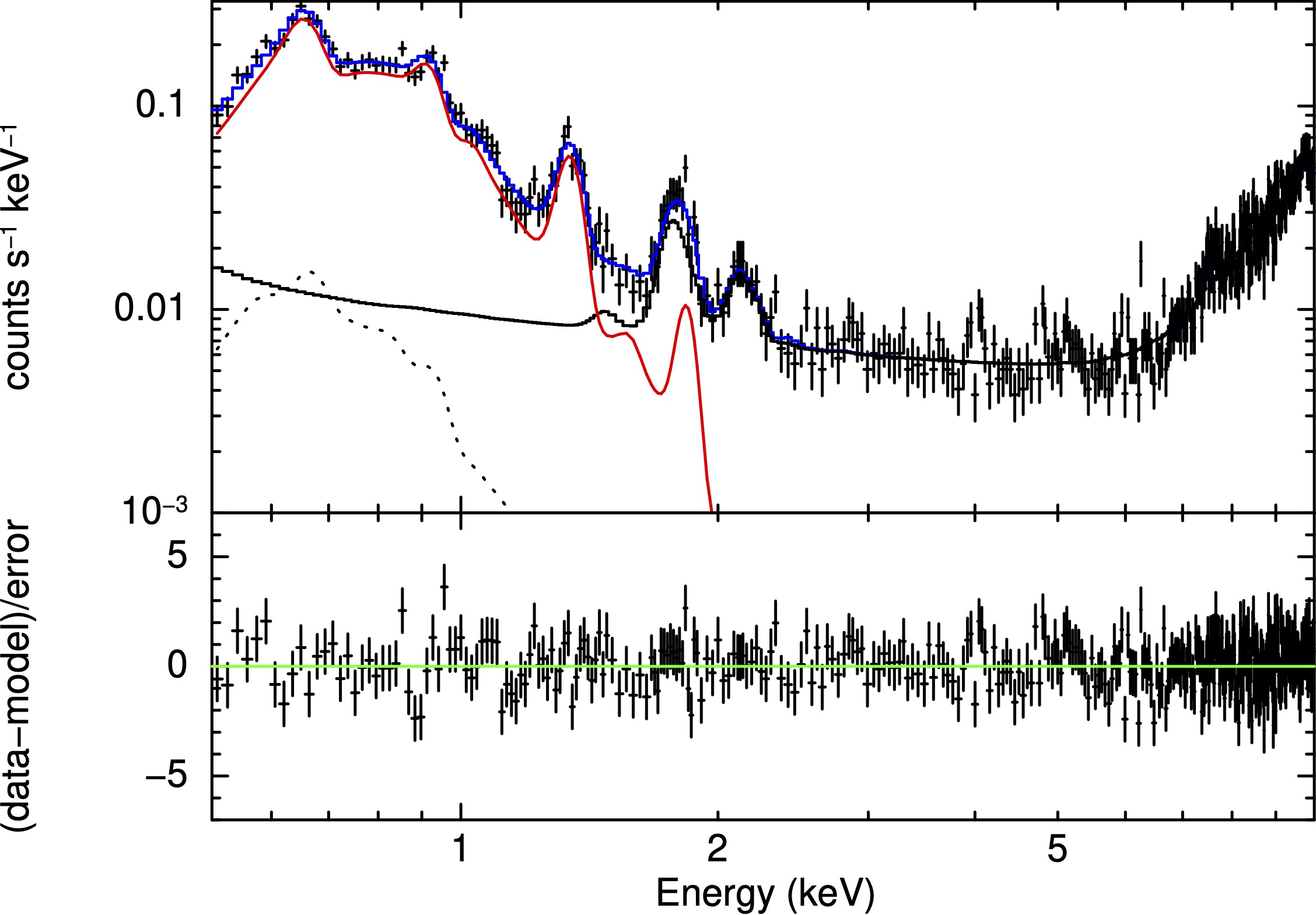}
    \caption{X-ray spectrum of the Mg-rich ejecta of J0550--6823 and the best-fit model with $\tau=1.0\times 10^{12} \rm ~s ~cm^{-3}$ (top panel). The red line represents the best-fit plasma model, which includes the effect of photon absorption ($\rm phabs\times vnei$). The black line represents the background spectrum generated by {\tt "mkacispback"}. The black dashed line represents the sky background model. The blue line represents the sum of all model components. The bottom panel represents the residuals of the fit.}
    \label{taufigure}
\end{figure}

In bright SNR observations, subtracting the background spectrum from a nearby blank region is sufficient for spectral analysis. However, in the case of faint and diffuse sources like J0550--6823, the contribution of the background emission becomes significant compared to the source signal, leading to a lower signal-to-noise ratio. As a result, the derived spectral properties can be sensitive to the choice of the background extraction region. In particular, the particle-induced background is dominant in this observation (see Figures \ref{background2} and \ref{taufigure}), and its contribution also varies within the CCD chip. To address this issue, we utilize the Chandra ACIS tool {\tt "mkacispback"}, which models the particle-induced background based on detector coordinates and time, achieving residuals within $\leq 10\%$ \citep{2021A&A...655A.116S, suzuki_2021_17210106}. By accounting for both spatial and temporal variations of the background, this tool enables more accurate modeling of the background spectrum. Thus, we constructed a spectral model for the particle-induced background within the source region (the ellipse region in Figure \ref{J0550region2}) using this tool in our study.

We modeled the sky background emission which was considered foreground emission using the spectrum taken from the rectangle region in Figure~\ref{J0550region2}. The background spectrum was fitted with an absorbed plasma emission model ($ \rm phabs \times apec$ in Xspec) as shown in Figure~\ref{background2}, while the particle-induced background was simultaneously modeled.
We found that a low-temperature plasma model with collisional ionization equilibrium ($kT_e$ = 0.22 keV) reproduces the spectrum well. We assumed that this emission is uniformly present within the field of view of this observation and incorporated it as the sky background model in the subsequent analysis.

After modeling the sky and particle-induced background, we use these for the spectral analysis of J0550--6823. The normalization of the sky background model was scaled by the ratio of the background to source extraction areas and applied to the source region of J0550--6823. This allowed us to incorporate the  background properties into the spectral analysis of the source. We fitted the X-ray spectrum with an absorbed foreground emission and non-equilibrium ionization plasma model ($\rm phabs \times (apec+vnei)$ in Xspec), as shown in Figure~\ref{taufigure}.

The best-fit parameters are summarized in Table~\ref{taubestfit_transpose}. As a result, we confirmed a Mg-rich composition in the remnant, where the Mg abundance is (Mg/H)/(Mg/H)$_\odot\gtrsim1$ even in different fitting conditions. In particular, when compared with Ne, which is produced in the same region as Mg, the results indicate that the Mg abundance is more than twice the solar ratio. Here, to fit the spectrum, the elemental abundances were fixed at 0.3 times the solar values, representative of the LMC metallicity \citep{2023ApJ...950...74S}, for all elements except for He (set to 1.0 solar), S (tied to the abundance of Si), and those explicitly listed in Table~\ref{taubestfit_transpose}, which were allowed to vary freely. Following \citet{2023ApJ...950...74S}, the absorbing column density was fixed at $N_{\rm H} = 0.16 \times 10^{22} \,{\rm cm^{-2}}$ for all spectral fits. Even if we assume a redshift equivalent to the velocity of the LMC \citep{2002AJ....124.2639V}, the conclusion remains unchanged. Uncertainties for each parameter are quoted at the 1$\sigma$ confidence level and account for possible variations in the other free parameters. Due to the limited photon statistics, it is difficult to determine the ionization state from spectral fitting; however, when treating the ionization timescale $\tau$ [$\mathrm{s \cdot cm^{-3}}$] as a free parameter, we obtained values exceeding $1 \times 10^{13}$ cm$^{-3}$ s. In the analysis, the ionization timescale was varied in four steps across the model grid to investigate its influence on the results. We here assume that the timescale of $1\times10^{12}~[\rm s\cdot cm^{-3}]$ is consistent with being ionization equilibrium \citep{2010ApJ...718..583S}, and the result of this ionization timescale shows the lowest $\chi^2$ value. Thus, we use it as the best-fit result. The Si abundance slightly varies depending on the difference in the ionization timescale. On the other hand, a low ionization age of $\lesssim 10^{11}$ cm s$^{-1}$ would not be reasonable for this old core-collapse SNR \citep[][; see also section \ref{origin}]{2014ApJ...785L..27Y}. 
The Si line at $\sim$1.8 keV is composed of the emissions from the remnant and background. Without the model emission from highly ionized Si, we see residuals around the Si line structure. This allows us to estimate the Si abundance of the remnant, although higher statistics or spectral resolutions are needed to robustly separate these two.

In young supernova remnants, both shocked interstellar medium (ISM) components and ejecta components heated by the reverse shock are present. However, in older remnants like the one studied here, both plasma components are likely to have cooled and mixed, making them difficult to distinguish through spectral fitting. Nevertheless, since multi-temperature structures can affect the determination of abundance ratios, we investigated their impact. We tested the addition of a low-temperature (0.2 keV) ISM component by adding an {\tt apec} model and found that the Ne/Mg ratio increased by 18\%, while the Si/Mg ratio decreased by 28\%. This change arises from a shift in the plasma temperature of the ejecta component, as a higher temperature increases the emissivity of the Si He$\alpha$ line. Therefore, such multi-temperature effects may represent a source of systematic uncertainty in our analysis. On the other hand, the high Mg abundance is robust even when accounting for this effect. 

\section{discussion}
Our analysis of Chandra ACIS data for J0550--6823 confirms that this remnant exhibits a Ne/Mg mass ratio of $\approx$1, suggesting an Mg-rich (Ne-poor) composition (Table~\ref{taubestfit_transpose}). This ratio is significantly different from those observed in typical core-collapse SNRs and from the average composition found in the LMC of $\approx$3. These characteristics suggest that the progenitor of J0550--6823 may have experienced a late-stage active shell burning event shortly before explosion \citep{2024ApJ...970....4M,2025ApJ...984..185S}. In the following discussion, we examine this interpretation from two perspectives: (1) by comparing the observed abundance ratios with predictions from one-dimensional pre-supernova models to evaluate the plausibility of a late-stage active shell burning origin, and (2) by considering the broader implications for the stellar population of the LMC, where we estimate the fraction of massive stars that may undergo similar evolutionary processes.

\subsection{The origin of J0550--6823}\label{origin}

\begin{figure*}[htbp]
    \centering
    \includegraphics[bb=0 0 1600 1200,width=\linewidth]{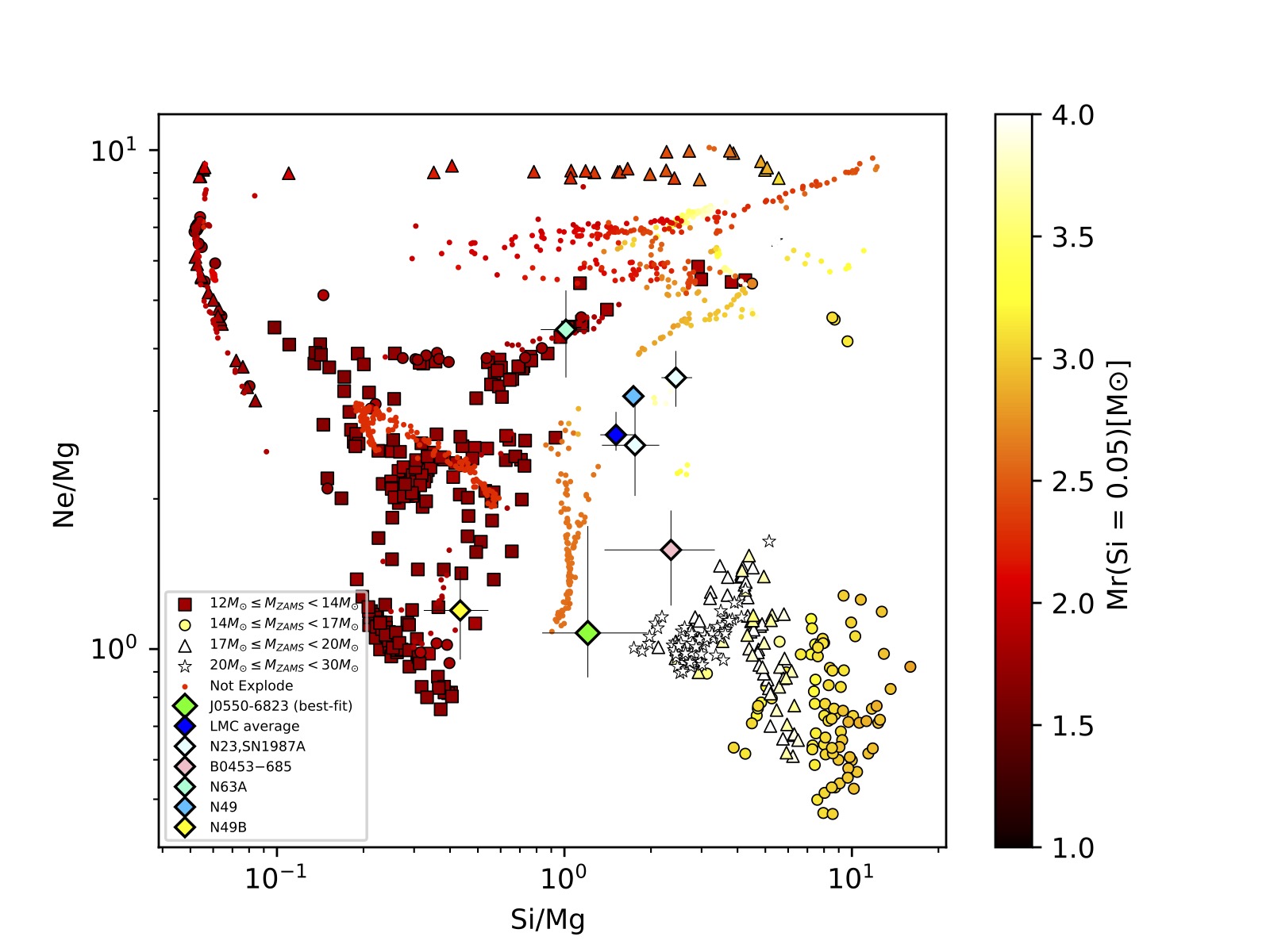}
    \caption{Si/Mg-Ne/Mg mass ratios of pre-supernova models and those of J0550--6823 (green plot). The green diamond marker shows the result with $\tau =1.00 \times 10^{12}\,[\rm s\cdot cm^{-3}]$ that we use as the best-fit result (See Table \ref{taubestfit_transpose}). The colored data points are derived from 1D pre-supernova models, with different markers used to represent each progenitor mass range \citep{2018ApJ...860...93S}. The radius of the Si-rich layer in the progenitor, $M_r$(Si=0.05), is shown in color. Using BH-NS separation curve for the w18.0 calibration that do not explode are identified and shown as colored dots \citep{2016ApJ...818..124E}. The SNRs of the LMC for which previous studies have shown the composition of Si , Ne, and Mg abundances are also plotted on the figure; N63A is from mean abundance of \cite{2003ApJ...583..260W}, B0453-685 is from vpshock model of \cite{2012A&A...543A.154H}, SN1987A is from vpshock W00 of \cite{2008ApJ...676..361H}, N49 is from \cite{2015ApJ...808...77U}, and N23 and N49B obtained values from \cite{2016A&A...585A.162M}.  The LMC average mass ratio is shown in blue marker \citep{2016AJ....151..161S}. 
    }
    \label{sukhboldplot}
\end{figure*}

To investigate the origin of J0550–6823’s Mg‐rich composition, we compare its observed Ne/Mg and Si/Mg mass ratios to a set of one‐dimensional pre–supernova models in Figure \ref{sukhboldplot} \citep{2018ApJ...860...93S}. Here, we introduce a simple diagnostic we call the ``Si radius'', $M_r(\mathrm{Si}=0.05)$: the maximum enclosed mass coordinate at which the Si mass fraction exceeds 0.05 (Figure~\ref{shellmerger}a). In models that undergo a shell merger, Si synthesized in the inner O‐burning shell is mixed outward, yielding a larger $M_r(\mathrm{Si}=0.05)$ (Figure \ref{shellmerger}). Therefore, stars that have experienced a shell merger are brighter (whiter) in color. The majority of the O-rich ejecta should have been synthesized during hydrostatic nucleosynthesis in stellar evolution, and its elemental composition is not so sensitive to explosive nucleosynthesis \citep[e.g.,][]{1995ApJS..101..181W,1996ApJ...460..408T}. Also, our previous research has discussed that the influence of explosive nucleosynthesis is sufficiently small \citep{2025ApJ...984..185S}. Thus, we just calculated these mass ratios in the O-rich layer of pre-supernova models with a high oxygen mass fraction $>$ 0.4, assuming that the effects of explosive nucleosynthesis can be ignored.

Shell-merger models occupy the lower-right region of the diagram, indicating Si- and Mg-rich compositions. In this diagram, the observed values of J0550–6823 are located at the left end of the distribution of the shell-merger model (see green data), which is not a perfect match. \cite{2023ApJ...950...74S} estimated the mass of the ejecta components by subtracting the ISM contributions (see Table~5 in their paper). They found that while the Si/Mg ratio in the ejecta remained unchanged even after the correction, the Ne/Mg ratio decreased, implying that the remnant was originally more Mg-rich. The fact that the Si/Mg mass ratio does not change even after the correction may suggest that there is little mixing of the O-burning products. Therefore, the expansion of the Ne shell inside the progenitor star (i.e., Ne-shell intrusion) may provide a better explanation for the observations. At the very least, the confirmed Mg-rich composition is likely indicative of some form of stellar activity.

The observation point is also positioned near the orange data points that are not expected to explode. According to \citet{2025ApJ...984..185S}, these are models of stars with initial masses in the $\sim$21--22 $M_\odot$ mass range, in which the Ne-burning shell has merged with the overlying convection zone. This model is a similar event to the Ne-shell intrusion seen in low-mass progenitors. However, in the Ertl criterion \citep{2016ApJ...818..124E}, these models are predicted to have difficulty exploding due to their high core compactness \citep[e.g.,][]{2020MNRAS.492.2578S}. The Ertl criterion is sometimes inconsistent with numerical simulations \citep{2022MNRAS.517..543W}, so these progenitors also retain the potential to explode. Whether or not these stars explode, active shell burning processes that promote Ne burning, such as shell mergers, are necessary to reproduce the Mg-rich composition in this remnant.

The observed elements may have been diluted by mixing with interstellar medium during the long time after the explosion. 
\cite{2023ApJ...950...74S} suggested that the age of this supernova remnant is $\approx$100 kyr using the same analysis based on a study by \cite{1974ApJ...188..501C}. We also estimated the age using the analytic solution by \cite{1988ApJ...334..252C}, which describes the evolution of a remnant in the radiative phase, and obtained an age of $\approx$160 kyr. Here, we used parameter values equivalent to those used in previous studies: a remnant radius of 40 pc, an ambient density of 0.4 cm$^{-3}$, and an explosion energy of 1$\times$10$^{51}$ erg. Thus, this remnant can be well explained as one that has swept up the ISM over a long period of time and is now in the radiative phase. In addition, the total mass of X-ray emitting gas is estimated to be $\sim$200--400 $\,M_{\odot}$ \citep{2023ApJ...950...74S}, which is significantly larger than the mass of a normal massive star. Therefore, if the ejecta heated by the reverse shock and the material gathered by the front shock are mixed during this long timescale, the composition ratio should be close to the average composition of the LMC. On the other hand, the observed Ne/Mg ratio is significantly lower than the LMC value (Figure~\ref{sukhboldplot}), supporting that it reflects the ejecta composition. Otherwise, the vicinity of the remnant must have a biased composition.

To further constrain the properties of the progenitor star, we need to determine the ionization state and Si abundance more precisely in the future. Our spectral analysis shows that the fitting with the highest $\tau$ has the lowest $\chi^2$ value (Table~\ref{taubestfit_transpose}), and this result is consistent with the general trend that many core-collapse SNRs show a large ionization age of $>10^{11}$ cm$^{-3}$ s from an early age ($<$ 1,000 years) \citep{2014ApJ...785L..27Y}, so the high $\tau$ would be reasonable for the remnant. Thus, we conclude that the ionization timescale near the ionization equilibrium is realistic in the case of J0550--6823. 

\subsection{The rate of Mg-rich events in the LMC}

Quantifying how often massive stars undergo shell mergers is essential for assessing their impact on supernova explosion conditions and the production of odd‐$Z$ elements.  The discovery of J0550--6823 as a second confirmed Mg‐rich SNR in the LMC after N49B provides the first empirical foundation for such estimates. In particular, because the LMC has a lower metallicity than the Milky Way, the emission from the ejecta can be observed for a longer time than in SNRs within the Milky Way, making the SNRs in the LMC more suitable for the systematic studies. Of the 59 confirmed remnants in the LMC, 31 are classified as core‐collapse events \citep{2016A&A...585A.162M}, but only seven have published Ne, Mg and Si abundance measurements suitable for identifying a shell‐merger signature (Figure~\ref{sukhboldplot}).

Within this subset, N49B and J0550--6823 exhibit low Ne/Mg ratios, supporting their classification as Mg-rich SNRs and suggesting that they may be remnants of progenitors that underwent shell merger events. A third object, B0453--685, also shows a comparatively low Ne/Mg ratio, even though it has not yet been formally classified as Mg-rich. This suggests it may represent an additional, as-yet unconfirmed merger candidate. Counting only N49B and J0550 yields an Mg-rich fraction of $\sim$28\% (2/7), which rises to $\sim$43\% (3/7) if B0453--685 is included as a provisional candidate. Extrapolating to all 31 core-collapse remnants gives a conservative lower limit of $\sim$6\% (2/31), or $\sim$10\% (3/31) including B0453--685.

It remains difficult to determine whether these features arise specifically from shell mergers or from other processes such as Ne-shell intrusion. Even among Mg-rich SNRs, not all may result from shell mergers. For example, if the merger is complete, Mg is produced by Ne burning but also destroyed by O burning \citep{2025A&A...698A.216R}. In this case, even if a shell merger occurred, there may be remnants in which Mg-rich ejecta are not observed. Therefore, the occurrence rate of Mg-rich SNRs does not necessarily reflect the true frequency of shell mergers. If all of these remnants are indeed related to shell mergers, then although the inferred fraction is still below the $\sim$50\% rate required to reproduce the Galactic odd-Z trends suggested by \citet{2018MNRAS.474L...1R}, these findings clearly demonstrate that shell mergers constitute a non-negligible channel for odd-$Z$ element synthesis.

Figure~\ref{sukhboldplot} illustrates how confirmed mergers (N49B, J0550--6823) occupy a distinct region of the Ne/Mg–Si/Mg plane, whereas typical remnants such as N63A lie in the standard locus.  As upcoming X‐ray missions and deeper spectral analyses expand the sample of well‐measured SNRs, these fraction estimates can be refined, yielding stronger constraints on the prevalence of shell mergers and their role in galactic chemical evolution. 

The current comparison with theoretical models is based on the massive star evolution calculations assuming the solar composition by \cite{2018ApJ...860...93S}.
In the low-metallicity environment of the LMC, stellar evolution can change and may lead to a different picture. On the other hand, activities such as shell mergers are thought to depend on the central carbon mass fraction ($X_C$) and the CO core mass \citep{2025A&A...698A.216R,2025A&A...695A..71L}, and effects such as binary interactions and the $^{12}\mathrm{C}(\alpha,\gamma)^{16}\mathrm{O}$ reaction rate may have a stronger impact than metallicity itself. In the future, more detailed theoretical modeling will be necessary to explain the fraction of the Mg-rich SNRs in the LMC.

\section{summary and conclusion}

We have reanalyzed deep Chandra observations of the LMC remnant J0550--6823 and confirmed that its ejecta exhibit an anomalously low Ne/Mg mass ratio of $\approx$1, far below the LMC average of $\approx$3. By fitting the X-ray spectrum, we confirmed this Mg‐rich (Ne‐poor) composition across a physically plausible range of ionization timescales. Comparing the observed Ne/Mg and Si/Mg ratios to a suite of one‐dimensional pre–supernova models \citep{2018ApJ...860...93S}, we find that only progenitors which have experienced vigorous late‐stage shell burning, such as a shell merger or Ne‐shell intrusion, can reproduce the observed abundance pattern. 

Taken together with the previously known Mg‐rich remnant N49B, our census puts at least 6\% (and perhaps up to $\sim$40\%) of core‐collapse supernovae in the LMC among Mg-rich SNRs and it is possible that they have experienced shell mergers or similar events. Such events not only reshape the stellar structure and can facilitate explosion by modifying the core compactness and seeding perturbations, but also boost the production of odd‐$Z$ elements (K, Sc, Cl, P) that are otherwise underproduced in standard yields.  If a comparable fraction of massive stars in other environments undergoes shell mergers, this mechanism may help resolve longstanding discrepancies in galactic chemical evolution models, particularly the underabundance of K and Sc in the models \citep[e.g.,][]{2006ApJ...653.1145K,2020ApJ...900..179K,2018MNRAS.476.3432P}.

In the future, more sensitive X‐ray spectroscopy \citep[e.g., with XRISM and Athena][]{2025PASJ..tmp...28T,2013arXiv1306.2307N} will help to measure the ionization age and Si abundance more precisely. In addition, systematic surveys of Ne/Mg ratios in older, faint SNRs can better quantify the true shell‐merger fraction.  On the theoretical front, three‐dimensional hydrodynamic simulations of marginal O–C shell interactions will illuminate the conditions under which mixing becomes nonlinear, and multi‐dimensional progenitor models will refine predictions for odd‐$Z$ and $p$‐process yields \citep[e.g.,][]{2018MNRAS.474L...1R}. Together, these efforts promise to transform Mg‐rich SNRs into a new observational window on the final convective life of massive stars.

\begin{acknowledgments}
We are grateful to Takaaki Tanaka and Natsuki Terano for valuable discussion. This work was supported by the Japan Society for the Promotion of Science (JSPS) KAKENHI grant No. JP23K13128. J.P.H., the George A. and Margaret M. Downsbrough Professor of Astrophysics, acknowledges the Downsbrough heirs and the estate of George Atha Downsbrough for their support. His Chandra research on supernova remnants was additionally supported by SAO award number G01-22058X. We thank the anonymous referee for their careful review and constructive comments, which improved the clarity and quality of this manuscript.
\end{acknowledgments}

\vspace{5mm}

\bibliography{sample631}{}
\bibliographystyle{aasjournalv7}

\end{document}